\documentclass[useAMS,usenatbib]{mn2e}
\usepackage[dvips]{graphicx}

\title[First images of 6.7-GHz methanol masers in DR21(OH) and DR21(OH)N]
      {First images of 6.7-GHz methanol masers in DR21(OH) and DR21(OH)N}
\author[L.~Harvey-Smith, R.~Soria-Ruiz, A. Duarte-Cabral and
      R.J.~Cohen.]{L.~Harvey-Smith$^{1,}$$^{2}$\thanks{E-mail: lhs@usyd.edu.au}, R.~Soria-Ruiz$^{1}$,
      A. Duarte-Cabral$^{1,}$$^{3,}$$^{5}$ and
      R.J.~Cohen$^{4}$\footnotemark[1]\thanks{Deceased}\\
$^{1}$Joint Institute for VLBI in Europe, Postbus 2, 7990 AA Dwingeloo, The Netherlands.\\
$^{2}$School of Physics, University of Sydney, 2006 NSW, Australia.\\
$^{3}$Departamento de F\'{i}sica da Faculdade de Ci\^{e}ncias da
      Universidade do Porto, Rua do Campo Alegre 687, 4169 - 007 Porto, Portugal.\\
$^{4}$University of Manchester, Jodrell Bank Observatory, Macclesfield,
      Cheshire, SK11 9DL, United Kingdom.\\
$^{5}$Jodrell Bank Centre for Astrophysics, University of Manchester, Alan Turing Building Oxford Road, Manchester, M13 9PL, United Kingdom.}

\begin{document}

\date{\today}

\pagerange{\pageref{firstpage}--\pageref{lastpage}} \pubyear{2007}

\maketitle

\label{firstpage}

\begin{abstract}
The first images of 6.7-GHz methanol masers in the massive star-forming regions DR21(OH) 
and DR21(OH)N are presented. By measuring the shapes, radial velocities and polarization properties of 
these masers it is possible to map out the structure, kinematics and magnetic 
fields in the molecular gas that surrounds newly-formed massive stars. The intrinsic angular resolution of the 
observations was 43 mas ($\sim$~100 AU at the distance of DR21), but structures far smaller than
this were revealed by employing a non-standard mapping technique. By plotting the positions of 
the Gaussian-fitted maser emission centroids in each velocity channel, the internal velocity gradients of 
the masers were investigated at very high spectral and spatial resolution. This technique was used in an attempt 
to identify the physical structure (e.g. disc, outflow, shock) associated with the methanol masers. Two distinct 
star-forming centres were identified. In DR21(OH) the masers had a linear morphology, and the 
individual maser spots each displayed an internal velocity gradient in the same direction as the large-scale structure. 
They were detected at the same position as the OH 1.7-GHz ground-state masers, close to the centre of an outflow traced 
by CO and class I methanol masers. The shape and velocity gradients of the masers suggests that they probably delineate a shock. 
In DR21(OH)N the methanol masers trace an arc with a double-peaked profile and a complex velocity gradient. This 
velocity gradient closely resembles that of a Keplerian disc. The masers in the arc are 4.5$\%$ linearly polarized, 
with a polarization angle that indicates that the magnetic field direction is roughly perpendicular to the large-scale magnetic field 
in the region (indicated by lower angular resolution measurements of the CO and dust polarization). The origin and nature of these maser 
structures is considered within the context of what is already known about the region. The suitability of channel-by-channel centroid 
mapping is discussed as an improved and viable means to maximise the information gained from the data.  

\end{abstract}


\section{Introduction} 

The DR21 star-forming complex is a large and highly active region, containing
many sites of recent and ongoing star-formation as evidenced by
observations of atomic and molecular gas and dust at a wide range of
frequencies. It lies in the Cygnus X region of our Galaxy at a
kinematical distance estimated to be between 2 and 3 kpc (Dickel \&
Wendker 1978; Campbell et al. 1982; Piepenbrink \& Wendker 1988;
Odenwald \& Schwartz 1993). DR21 contains many sub-regions denoted by
positions of Infra-Red Sources (IRSs), Far Infra-Red sources (FIRs) or
Extremely Reddened Objects (EROs), which are interpreted as sites of ongoing
star-formation (e.g. Kumar et al. 2007; Jackob et al. 2007). This paper
will describe the results of Multi-Element Radio-Linked Interferometer 
Network (MERLIN) observations of class-II methanol masers
at 6.7-GHz in two regions of active massive star-formation within DR21.

The aim of this work is to understand the role of outflows, circumstellar discs and 
magnetic fields in the formation of massive stars. This can be achieved by studying 
the 6.7-GHz methanol masers at high spatial and spectral resolution. It is possible to 
effectively enhance the resolving power of MERLIN by measuring the centroid positions of the maser emission in each spectral 
channel of the maser line. This, coupled with linear and/or circular polarization 
information on the masers allows us to form a 3-dimensional $+$ magnetic field picture 
of the physical region immediately surrounding emerging massive stars. 
 
This work is part of an on-going study of several regions of massive
star-formation that produce bright OH and methanol masers. Data from phase-referenced observations at 1.7, 4.7, 6.0 and
6.7-GHz are also available to the authors. This sample covers all the OH and methanol maser lines that
can be observed within the working frequency range of MERLIN.  Parts of this
survey have already been published (Harvey-Smith \& Cohen 2005;
Nammahachak et al. 2006; Green et al. 2007) but the mapping of several
regions is on-going. Most of the observations in the multi-frequency
study were correlated in full polarization mode, and it is therefore possible to
measure the properties of magnetic fields in these regions, as well as 
the Faraday rotation (e.g. Vlemmings, Harvey-Smith \& Cohen 2006). The goal of this survey is to 
map all the maser lines listed above in full polarizations, thus forming an in-depth picture of several regions of massive star-formation from 
which to draw specific conclusions about the relationship between masers, magnetic fields and stellar accretion. 

\section{Observations}

The 5$_1$--6$_0$ A$^+$ 6668.518 MHz methanol emission from DR21(OH)
   (also known as W75S or W75S(OH)) and DR21(OH)N (also called W75) was observed between
   January 12th and 14th 2005 using five telescopes of MERLIN: the Mark
   II telescope at Jodrell Bank and the outstation telescopes at Cambridge, Darnhall,
	Knockin and Pickmere. The longest baseline was 218~km, giving a
	minimum fringe spacing of 43~mas at 6.7~GHz. A spectral
	bandwidth of 0.5 MHz was employed for the line observations
	(narrow-band mode) and this was split into 512 frequency channels. The data were correlated in full polarization
	mode. The radial velocity range was 22.5 km~s$^{-1}$, centred on a velocity of 5 km~s$^{-1}$. The velocity resolution was 0.0439 km~s$^{-1}$. Here and elsewhere radial velocities are given relative to the local standard of rest.
	For observations of the phase-calibrator source 2005+403, a 16-MHz
	bandwidth was split into 16 channels (wide-band mode) and then
	averaged into a single channel. Every 6 or 7~minutes on DR21(OH) in narrow-band mode was followed by a 2~minute period on the
	phase-calibrator in the wide-band mode. The same procedure was
	followed for DR21(OH)N. At 6.7 GHz, the total time spent on each
	target source was 2 hours. In addition, two longer periods (a total of 4
	hours) were spent on the bandpass calibrator source 3C84, switching between wide-band and narrow-band
	modes (after a 3 minute observation period in each mode). Another two
	extended periods of a total of 4 hours were spent on the primary flux and
	polarization-angle calibrator 3C286 (again switching from wide-band
	and narrow-band modes every 3 minutes). 

Instrumental feed polarization was determined using {\sc aips} task
{\sc pcal}. After determining the phase offset between the RCP and LCP
signals, self-calibration was performed on a strong and compact maser
feature using the left circular polarization image. The calibration
results were then applied to both polarizations. A sample of 5 
maser features were tested, but no statistically significant shift 
in the positions of the masers was found before and after self-calibration had been applied. 
The absolute polarization angle of the masers was determined using 3C286, which 
has a known polarization position angle of 33$^\circ$. 

A region of 512 $\times$ 512 pixels (7$\farcs$68 x 7$\farcs$68) was
imaged in Stokes I,Q,U and V with a pixel size of 15 mas. A circular
restoring beam of 50 mas was used in the {\sc aips} task {\sc
  imagr}. In order to improve the signal-to-noise ratio of the images,
self-calibration was applied to the Stokes-\emph{I} image before the
positions and velocities of the masers was calculated. 

A careful search of the MERLIN 6.7 GHz Stokes-\emph{I} image led to the identification
of all the methanol maser flux in DR21(OH) and DR21(OH)N within the
calibration errors. This was in the form of compact, but elongated,
methanol masers. Two-dimensional Gaussian components were fitted to the
images using the {\sc aips} routine {\sc orfit} and the positions and
peak velocities of the maser components in each channel were found. 

For this study, a maser was deemed to exist where an emission feature with a minimum
flux density of 5 times the RMS noise level in that channel appeared in
at least six adjacent velocity channels. It became clear however, that rather
than being simple maser spots, the vast majority of maser components
had significant internal structure, meaning that the automatic
assignment of mean positions and velocities to maser spots would lead
to a loss of information. The data are therefore presented both as
maser spots and as Gaussian-fitted centroid positions of the maser
emission in individual channels. As the following sections will show, this technique has led to much greater understanding of
the masers, including their morphologies and radial velocity profiles.

\section{Results}

\subsection{Overview of results}

Two sites of 6.7-GHz methanol emission were detected within the DR21
star-forming complex. This was in the form of a group of five maser spots in DR21(OH)
and three maser spots in DR21(OH)N. Two of the spots in DR21(OH)N seem
to make up a single physical structure which is producing linearly
polarized methanol masers. No existing maps of 6.7-GHz methanol maser
emission in these regions have been found and so all the masers in this paper are thought to be new
discoveries. Table 1 lists the flux weighted mean positions, peak flux densities,
brightness temperatures, velocities and linewidths of the maser spots in
DR21(OH)N and DR21(OH). Figure 1 shows the spectra measured at 6.7 GHz. 
There are clearly two strong maser peaks in DR21(OH)N, whereas DR21(OH) shows a single, weaker and more
extended peak.

\begin{table*}
\begin{center}
\caption[Parameters of 6.7-GHz methanol masers in DR21(OH)N and
  DR21(OH).]{\label{methanoltable}Parameters of 6.7-GHz methanol masers in DR21(OH) and DR21(OH)N.}
\begin{tabular}{ccccccc}
\hline
 Maser Spot & R.A. (J2000)  & Dec. (J2000)   & Peak Flux Density & Lower  Limit to & Velocity & $\Delta V_{1/2}$ \\
 Number    & \hspace{0.3cm}(20$^h$39$^m~~^s$) &  (42\degr24\arcmin~~\arcsec)  & (Jy beam$^{-1}$) & Peak $T_b$ (K) & (km~s$^{-1}$) & (km~s$^{-1}$)   \\
\hline
1 & 0.377  & 37.145  & 9.280 & 1.48$\times$10$^{8}$ & 4.555 & 0.38\\
2 & 0.379  & 37.150  & 4.801 & 7.64$\times$10$^{7}$ & 3.559 & 0.25\\
3 & 0.373  & 37.182  & 0.253 & 4.03$\times$10$^{6}$ & 3.199 & 0.12\\
\hline
  & \hspace{0.3cm}(20$^h$39$^m~~^s$) & (42\degr22\arcmin~~\arcsec)  & &  & &   \\
4 & 1.052 & 49.177 & 1.610 & 2.56$\times$10$^{7}$ & --2.800 & 0.28 \\
5 & 1.074 & 49.287 & 0.303 & 4.82$\times$10$^{6}$ & --3.109 & 0.20 \\
6 & 1.066 & 49.233 & 0.848 & 1.35$\times$10$^{7}$ & --3.133 & 0.20 \\
7 & 1.071 & 49.257 & 0.450 & 7.16$\times$10$^{6}$ & --3.401 & 0.13 \\
8 & 1.026 & 49.026 & 0.537 & 8.55$\times$10$^{6}$ & --3.887 & 0.25 \\
\hline
\end{tabular}
\flushleft
\hspace{0.8cm}
\label{posvelresults}
\end{center}
\end{table*}

\begin{figure}
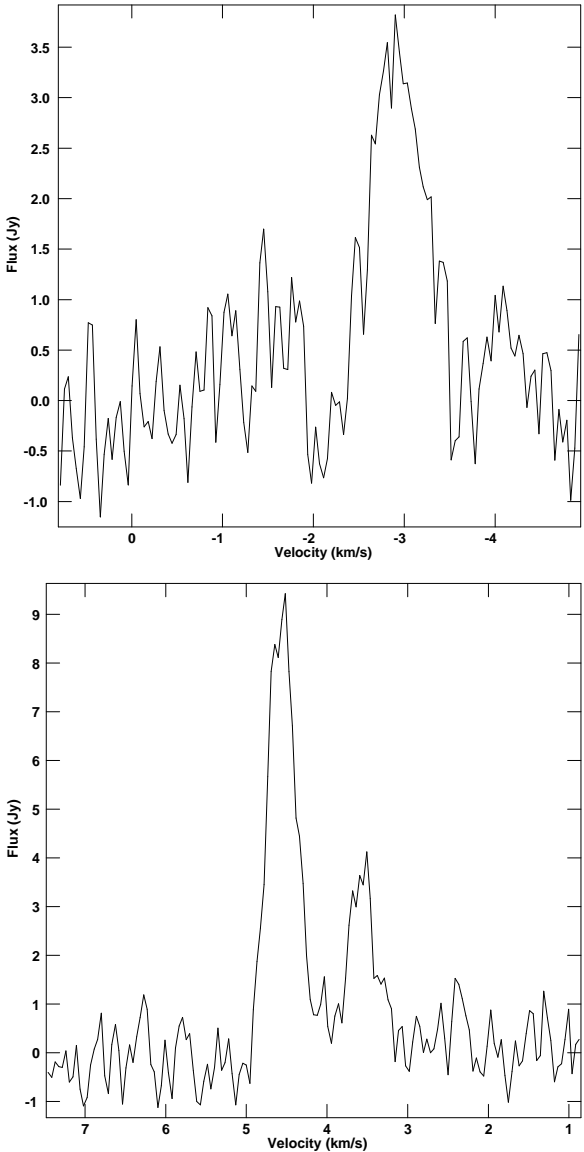

\begin{center}
\includegraphics[angle=-90, width=8 cm]{fig1a.ps}
\includegraphics[angle=-90, width=8 cm]{fig1b.ps}
\caption[Methanol maser spectra in DR21(OH) and DR21(OH)N]{Methanol
  6.7-GHz maser spectra in DR21(OH) (top) and DR21(OH)N (bottom).}
\end{center}
\end{figure} 

In the following sections, the properties of the masers found in each region are discussed in more detail.

\subsection{DR21(OH)}  

By plotting the positions of Gaussian-fitted centroids of the maser
emission channel-by-channel, the velocity gradients within DR21(OH) were mapped out in great detail. Figure 2 shows the positions of the 6.7-GHz maser emission above
5$\sigma_{RMS}$ in each spectral channel in DR21(OH). The elongation of
the maser emission is clear, with all the maser spots sharing a common
position angle in this region. It is interesting to note that each
maser spot is elongated in the same plane as the large-scale
structure. Each spot has an internal velocity gradient with the
same orientation, although the region as a whole has no clear velocity
gradient. This strongly indicates that the linear arrangement of
methanol masers delineates neither a circumstellar disc nor an
outflow. One explanation could be that the molecular cloud as a whole
is rotating, and the masers are triggered by a shock propagating
through the cloud. Thus the individual maser groups have very similar
velocity gradients along their length. This model will be discussed further in Section 4.1.

Figure 3 (left) shows flux density contours of the 6.7-GHz maser emission in
DR21(OH), integrated between 2.6 and 4.1~km~s$^{-1}$. Figure 3 (right) shows the
velocity versus R.A. through the maser structure, integrated over all declinations
across the feature. This reveals that the linear structure is actually
split into three separate sections, marked A, B, and C, each sharing a
similar velocity gradient but having a separate location on the sky. 

\begin{figure*}
\begin{center}
\vspace{0.6 cm}
\includegraphics[angle=0, width=16 cm]{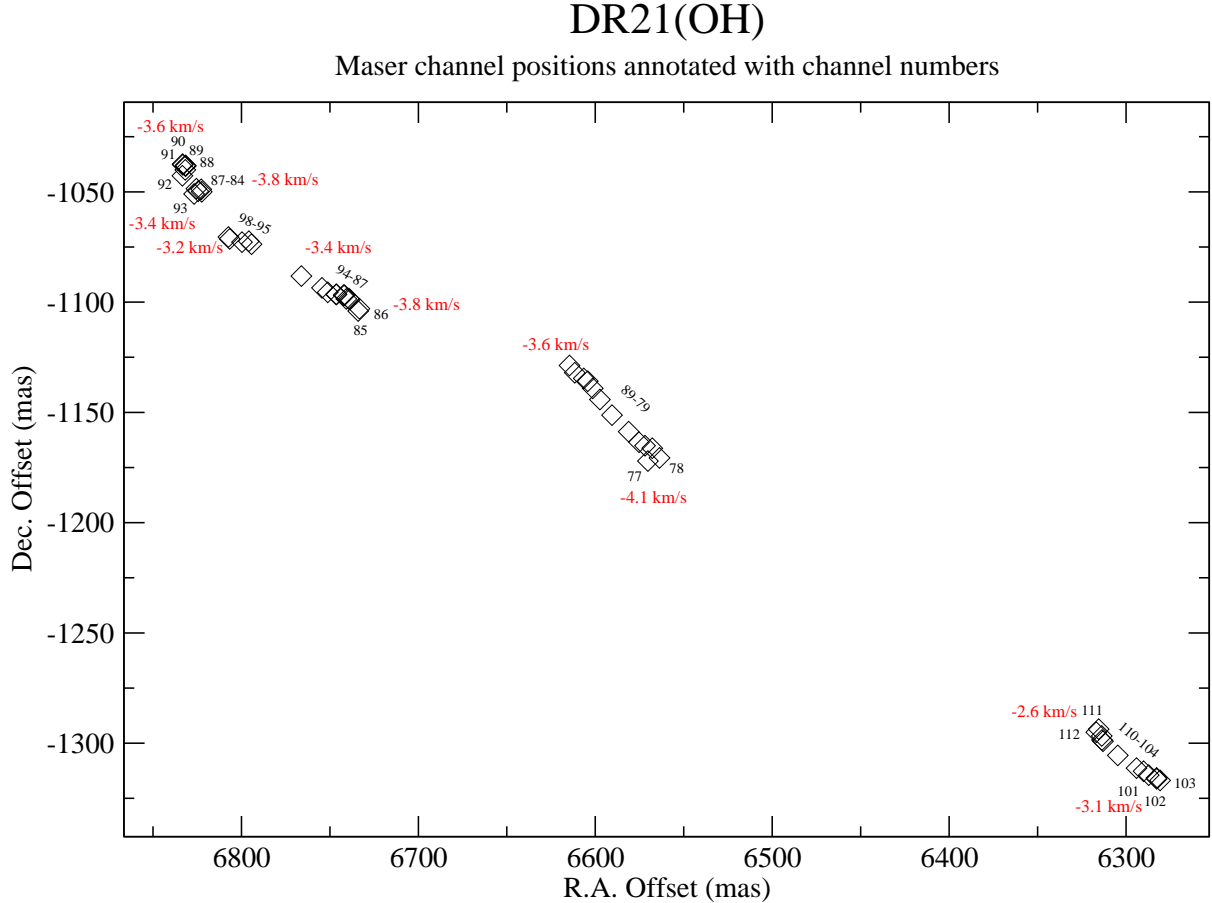}
\caption[Channel map of 6.7-GHz methanol masers in
  DR21(OH)]{Channel-by-channel positions of 6.7-GHz methanol masers in
  DR21(OH), annotated with channel numbers. R.A. and Dec. offsets are
  relative to the map centre at 20$^{\circ}$39'00$\farcs$4570 +42$^{h}$22$^{m}$50$\fs$330. The maser emission is clearly aligned along a linear
  structure, $\sim$0.7 arcseconds in length. Within each group of
  masers indicated on the map (i.e. within each maser spot), the
  redshifted velocity increases systematically towards the West,
  directly along the linear structure. However, there is no global
  velocity gradient across the region as a whole.}{\label{label here}}
\end{center}
\end{figure*} 

\begin{figure*}
\begin{center}
\includegraphics[angle=0, width=16 cm]{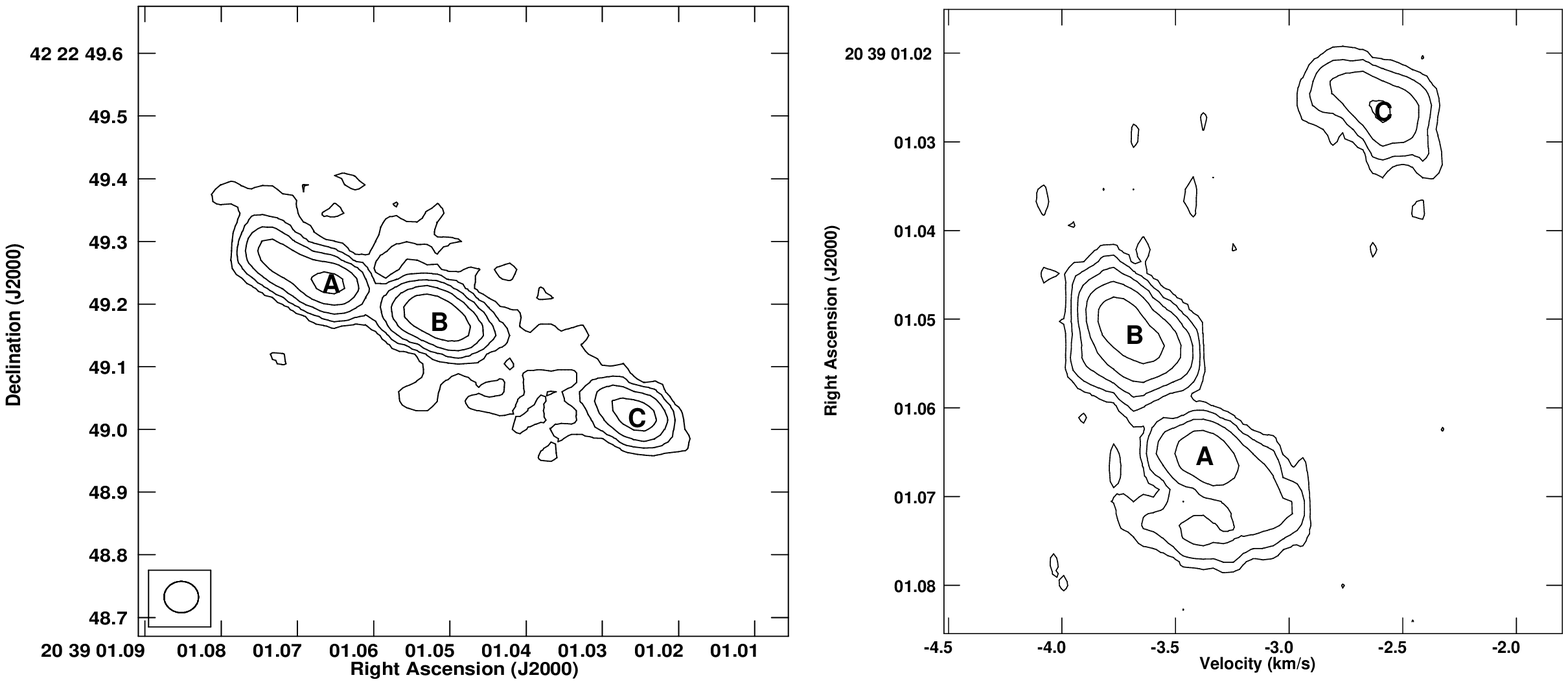}
\caption[Methanol maser filament in DR21(OH)]{Left: Methanol
  6.7-GHz maser flux in DR21(OH), integrated between 2.6 and 4.1 km~s$^{-1}$. The peak flux is 186 mJy/beam
  and the contour levels are 6 $\times$ (1, 2, 4, 8, 16, 32,
  64, 128, 256, 512, 1024) mJy~beam$^{-1}$. Right: Velocity--R.A. plot
  of the region. Sections A, B and C have similar velocity gradients,
although section C is moving towards us with a greater velocity than
sections A and B. The fact that the velocity gradient is
similar in each section of the filament suggests that this is caused by
a bulk rotation in the cloud.}
\end{center}
\end{figure*}

\subsection{DR21(OH)N} 

Positions of the Gaussian-fitted centroids of the maser emission in
DR21(OH)N are plotted channel-by-channel in Figure 4. There are three maser `spots' by the
common definition (as described in the Introduction), however there appear to be only two separate
physical structures traced by maser emission, each of which has an
ordered velocity gradient. These maser features will be referred to as components 1 and 2. 

Component 1 has a fascinating arc structure, with velocities
ranging from 3.2 to 4.9 km~s$^{-1}$ at either end. This feature has an
internal velocity gradient that is complex but very well sampled. Figure 5 shows this velocity
gradient along the major (East--West) axis across the 39 spectral channels showing
maser emission to a level of 5$\sigma_{RMS}$ in this feature. Its shape
resembles the Keplerian rotation curve of an edge-on disc, however
there are clearly some deviations from this behaviour. It has a linear velocity gradient
through the central region near channels 84 and 85, where the radial
velocity changes very little across the $\sim$20 mas width of the
feature. Then there are the two Keplerian-like portions of the
structure. 

Figure 6 shows the polarization properties of the masers in
DR21(OH)N. There two large peaks in the spectrum of linear polarization;
these occur in the West and East of the maser component 1. These maser spots are 4.9$\%$ and 4.5$\%$ linearly polarized
respectively. Both peaks share a linear polarization angle of
approximately --40$^{\circ}$, although there appears to be a linear gradient in
polarization angle, with the orientation changing by 20 degrees across
the feature. Assuming that the magnetic field direction is
perpendicular to the linear polarization vectors (Vlemmings et al. 2006), the mean orientation of the magnetic field
vectors in DR21(OH)N is +50$^{\circ}$ East of North.  

Component 2 is simple in morphology, unpolarized, and has a linear
velocity gradient across the feature between 3.1 and 3.3
km~s$^{-1}$. In the following sections, these results will be discussed in
the context of what is already known about the regions. 

\begin{figure*}
\begin{center}
\vspace{0.4cm}
\includegraphics[angle=0, width=14 cm]{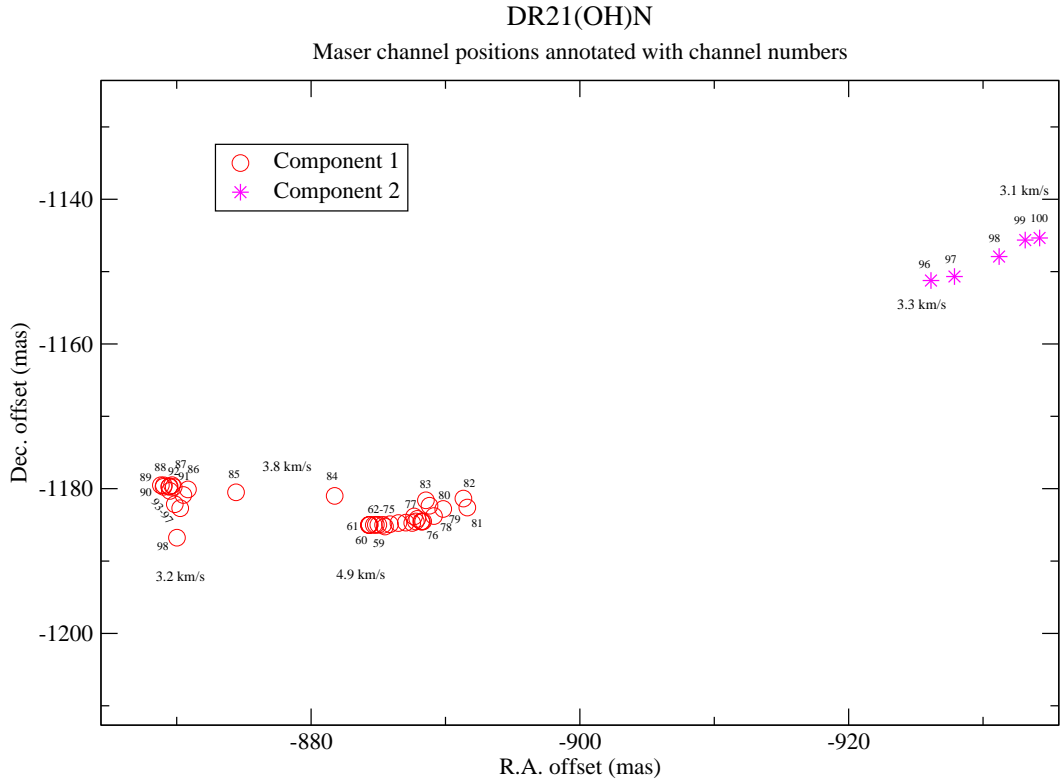}
\caption[]{Map of 6.7-GHz methanol masers in DR21(OH)N. Symbols
  represent the Gaussian fitted centres of emission in each spectral
  channel with signal above 5$\sigma$ times the RMS noise level in that
  channel. R.A. and Dec. offsets are
  relative to the map centre at 20$^{\circ}$39'00$\farcs$4570 +42$^{h}$24$^{m}$38$\fs$330. There are three separate maser spots, although two of these
  (spot numbers 1 and 2 in Table 1) appear to be physically
  associated. Here they are referred to collectively as component 1.}{\label{label here}}
\end{center}
\end{figure*} 

\begin{figure*}
\begin{center}
\vspace{0.4cm}
\includegraphics[angle=-90, width=14 cm]{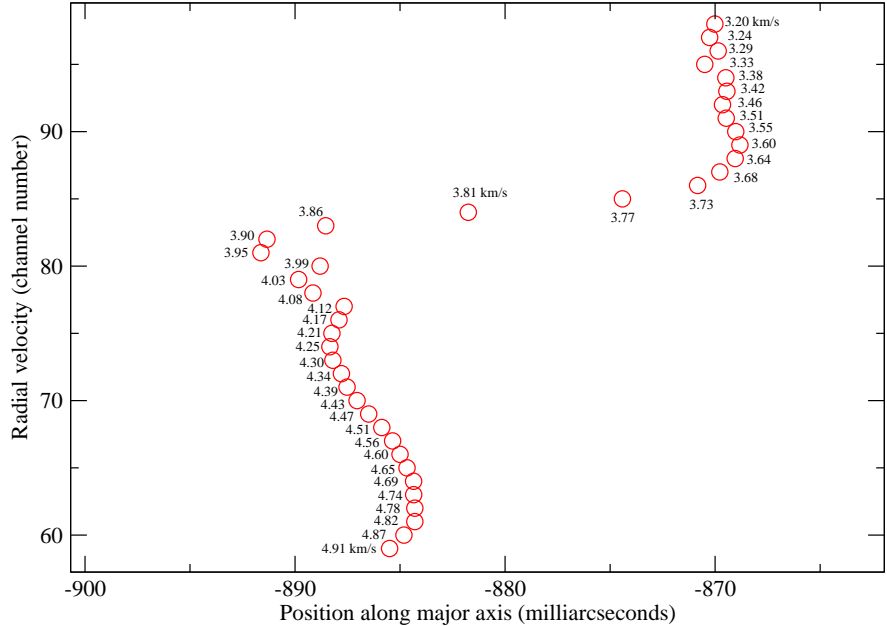}
\caption[]{The velocity gradient of component 1 along its major
  (East--West) axis. Each symbol represents a single spectral channel and is labelled with the central velocity in km~s$^{-1}$. The structure is well sampled, with 39 velocity
  channels having a singal greater than 5$\sigma$ times the RMS noise level. The position along the major axis is measured in
  milliarcseconds, relative to the pointing centre of the image.  }{\label{label here}}
\end{center}
\end{figure*} 

\begin{figure}
\begin{center}
\includegraphics[angle=-90, width=8 cm]{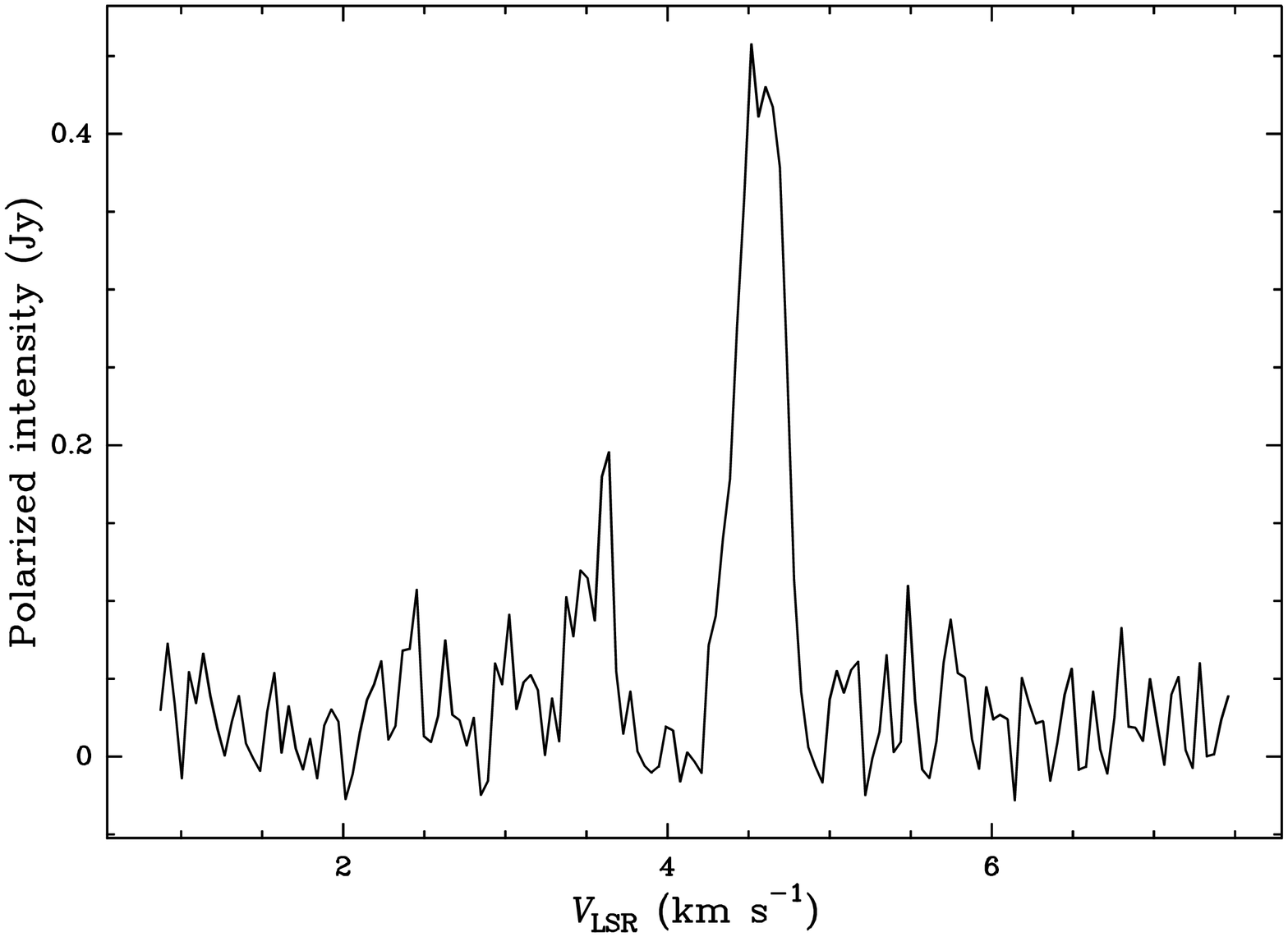}
\includegraphics[angle=-90, width=8 cm]{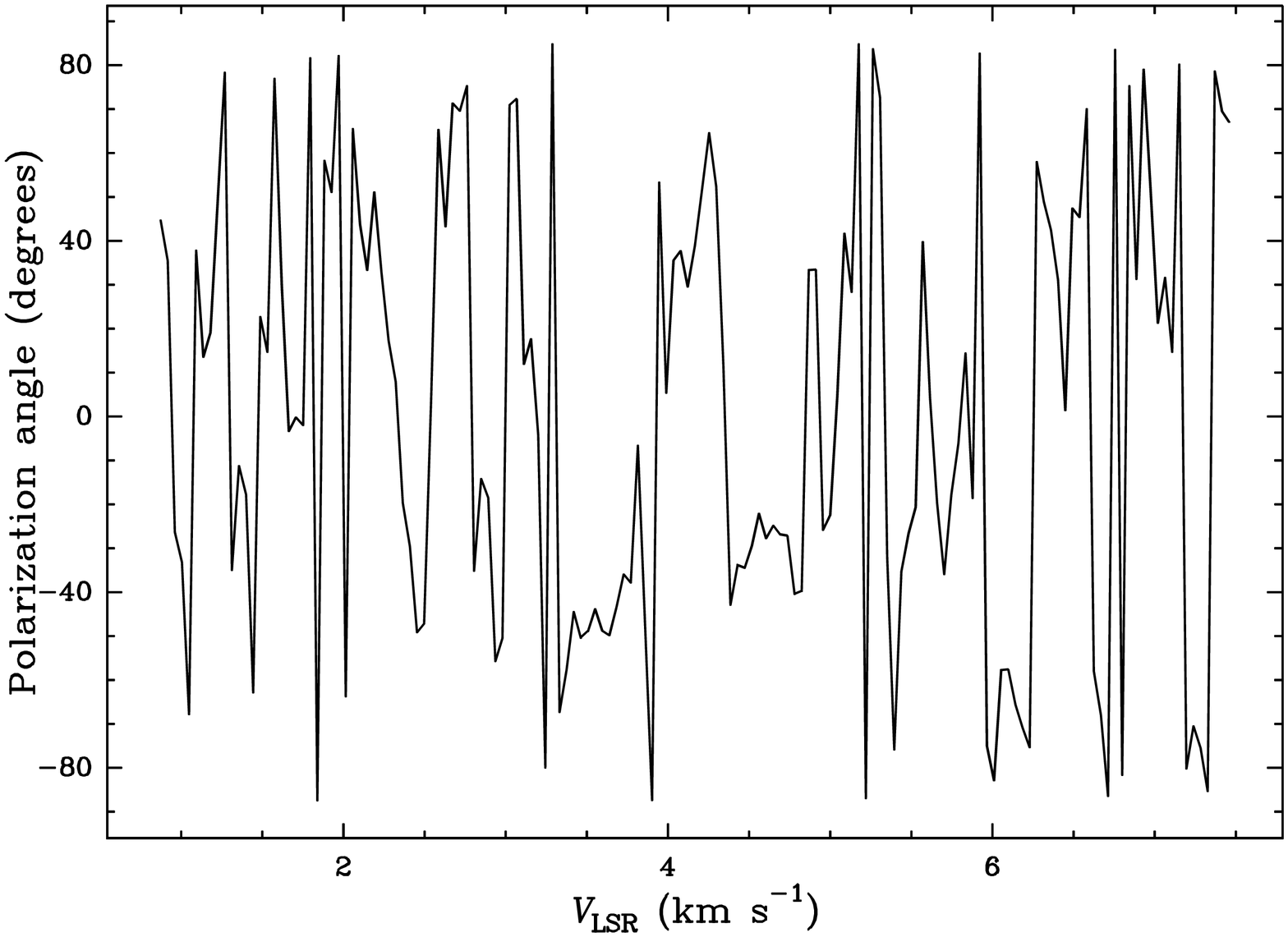}
\caption[]{The polarization properties of DR21(OH)N. The linear
  polarization spectrum (top) shows a large and a smaller peak, which
  correspond to the West and East regions of component 1,
  respectively. The polarization angle plot (bottom) shows two coherent
  sections at the velocities of the peaks in polarized intensity, with
  a polarization position angle of around --40$^{\circ}$. There appears
  to be a linear gradient in the polarization position angle across
  component 1.} {\label{label here}}
\end{center}
\end{figure}

\section{Discussion}

\subsection{The molecular environment of DR21(OH)}

DR21(OH) appears to be composed of two deeply embedded star-forming
cores, called MM1 and MM2, which are traced by 1.4 and 2.7~mm continuum
emission from dust and $J = 2 \rightarrow 1$ C$^{18}$O (Woody et
al. 1989, Padin et al. 1989, Magnum, Wooten \& Mundy 1991, Lai, Girart
\& Crutcher 2003). The cores are young, having no visible {\sc Hii} region and
therefore no continuum emission is seen from the region at centimeter
wavelengths. From measurements of the far infra-red luminosity, MM1 is
thought to contain a B0.5 V ZAMS star (Magnum, Wooten \& Mundy,
1992). The MM1 core also contains ground-state OH masers at 1.6 GHz
near the peak of the dust continuum emission (Norris et al. 1982; Fish
et al. 2005).  Next to MM1 is MM2, a dense and massive (480~$M_{\odot}$) condensation which, according to estimates of the
dust temperature, may contain at least one massive early-B type star
(Magnum et al. 1991). Interferometric observations of NH$3$ (1,1)
emission using the VLA resolved MM2 into two components, called MM2--A
and MM2--B (Magnum et al. 1992). 

Centred approximately on the MM1 and MM2 cores, there is a region of
extended methanol continuum emission which, towards its outer edges, contains class I methanol masers at 36 GHz 
(Haschick \& Baan 1989), 44 GHz (Haschick, Menten \& Baan 1990), 81 and
84 GHz (Batrla \& Menten 1988) and 95 GHz (Plambeck \& Menten
1990). Measurements of the methanol continuum, CO $J = 3-2$ emission
and the Class I methanol masers all suggest the presence of an outflow centred close to MM1/MM2 which is aligned approximately East--West (Lai et al. 2003; Vallee \& Fiege 2006).  

Perpendicular to this outflow is an extended region of CS molecular
emission, mapped by Chandler et al. (1993), which is part of the large-scale molecular ridge in DR21.
Water masers at 22 GHz have been observed to coincide with this CS line (Genzel \& Downes 1977; Magnum et al. 1992). The morphology of water masers detected in DR21(OH) are
not consistent with an outflow, rather they appear to have formed at the shock boundaries of
expanding stellar shells centred on the star-forming cores MM1, MM2--A and MM2--B.  

The magnetic field direction in DR21(OH) -- as elsewhere in the DR21 complex -- has been
determined to be perpendicular to the large-scale North--South finger of gas and
dust that connects the individual star-forming regions (Lai et al. 2003; Vallee \& Fiege 2006). Unfortunately, the methanol maser emission in DR21(OH) 
was unpolarized and therefore no additional information can be added on the magnetic field. 

\subsection{A filament of methanol masers in DR21(OH)}

A line of 6.7-GHz methanol masers was detected, that appears to be
associated with the OH masers at the central peak of MM1. This 1.4mm
continuum source is thought to house an embedded pre-main sequence B-type star.  
The maser filament in DR21(OH) is well sampled in position and velocity
space (see Figure 2). By studying the velocity gradients channel-by-channel, it becomes clear that the maser spots
each have independent but similar velocity gradients. There is no clear
systematic velocity gradient for the whole region however, which precludes the
possibility of this maser filament delineating a single disc or
outflow. The linear maser spots appear not to be tracing individual
circumstellar discs either, as the calculated masses of the central stars (assuming Keplerian
rotation) are small fractions of a solar mass
($\sim$10$^{-2}M_{\odot}$). Instead, the spatial morphology seems to
indicate that the masers were most likely to have been formed in a
shock passing through a turbulent or clumpy medium. 

Dodson, Ojha \& Ellingsen (2004) proposed a model to explain the high
prevalence lines or arcs within methanol masers. They described a
scheme in which the masers arise in planar shocks travelling
perpendicular to the line-of-sight. Their model fitted with their observations of velocity gradients within individual methanol maser spots
at 6.7 GHz that were perpendicular to the orientation of the lines or
arcs of masers. In contrast, the opposite appears to be true in DR21(OH). Here, the internal velocity gradients of the masers are parallel to the
direction of the linear extension. This could still be compatible with the passage of an edge-on shock if (1) the rotation velocity of the cloud as a whole were much greater than the velocity of the shock, in other words, there was a near-stationary density enhancement in the
rotating star-forming cloud, or (2) the masers were produced in a very
narrow region of the shock, so the motion of the shock itself is not
visible in the velocity gradient of the maser. 

\subsection{DR21(OH)N -- methanol masers in a protostellar disc?}

The 6.7-GHz methanol maser emission found near DR21(OH)N coincides with
the bright 8 $\mu$m source ERO3 within the 0.4 arcsecond positional
uncertainties of the \emph{Spitzer} IRAC survey (Marston et al. 2004; Davis et al. 2007).       
ERO3 has a spectral index (corrected for dust extinction) of 1.8 and it
is thought to be a very young massive protostar that has not yet
developed an {\sc Hii} region. It is therefore possible that component
1 represents a disc around the ERO3 protostar. The size of the disc traced by the methanol masers, assuming that the disc is seen edge-on 
and the masers are in a circular orbit, is approximately 60 AU. The velocity profile of the masers resembles the rotation curves (from observations and models)
of the black hole disc in the nucleus of Centaurus A (Marconi et al. 2001). It is likely that the departure from a standard Keplerian
rotation curve observed in DR21(OH)N is caused by a combination of factors; namely an inclination of the disc to our line-of-sight and a
difference between our selected East--West disc major axis and the true orientation. Existing MERLIN OH maser data are currently being analysed to test whether the feature is seen in OH and to confirm the direction (and possibly the strength) of the magnetic field in DR21(OH)N. A proposal has 
been accepted by the European VLBI Network to observe the DR21 region at 6.7-GHz in order to independently measure the positions and velocities of this disc-like feature at milliarcsecond resolution. Lastly, a quantitative rotational model is being developed in order to determine the nature of this very interesting 
and unusual maser feature (Harvey-Smith et al. \emph{in prep.}).

\subsection{Magnetic fields in DR21}

DR21(OH) is part of the much larger DR21 star-forming complex which is
crossed by a prominent North--South finger of dust, seen in thermal 850-$\mu$m
(e.g. Davis et al. 2007; Vallee \& Fiege 2006) and 1.3~mm continuum
emission (Lai et al. 2003). Comparisons of the CO and
dust polarization characteristics by Lai et al. (2003) show that the
magnetic field runs in roughly an E-W direction, perpendicular to the large-scale molecular
ridge. Curran et al. (2005) also confirmed this magnetic field orientation by measuring dust 
and CO polarization in DR21(OH). Whilst the angular resolution of these studies was far lower than the MERLIN 
observations reported here, it is nevertheless possible to undertake a comparison of the magnetic field properties
inferred from both the CO and dust emission (at $\sim$10'' scales) and the methanol masers (at sub-arcsecond scales). 
The aims of such a comparison are (i) to facilitate an understanding of the variations in magnetic fields inside giant molecular 
clouds and (ii) to gauge the importance of magnetic fields in the process of protostellar accretion.

There is very little published information on the magnetic field in the vicinity of our linearly polarized methanol masers in DR21(OH)N. 
The only polarimetric data available is in the paper of Vallee \& Fiege (2006). They measured the dust polarization at 850-$\mu$m using the James Clerk 
Maxwell Telescope and found, as in previous studies of DR21(OH)S and DR21(OH), that the inferred magnetic field direction is generally perpendicular to 
the North--South dust filament. Their polarimetric data had a angular resolution of 14'' and they binned squares of 3$\times$3 pixels together, giving 
images with a pixel spacing of 9''. In contrast, the MERLIN images of methanol masers from this paper have an intrinsic angular resolution of 50~mas and a pixel size of 15~mas. The dust polarization measurement closest in the sky to our polarized methanol masers gave a position angle of 172$^{\circ}$$\pm$13$^{\circ}$. This is not at all consistent with the methanol maser polarization position angle of 50$^{\circ}$, which was measured at both the East and West points of methanol maser component 1. 

So why does the magnetic field direction derived from observations of methanol masers disagree with previous measurements of CO and dust polarization by Vallee \& Fiege (2006)? There is more than one possible explanation for this discrepancy. Firstly, the interpretation of the magnetic field direction from the measurements of CO or maser polarization angle may be incorrect. Goldreich, Keeley \& Kwan (1973) showed quantitatively that, in the case that there is no circular polarization, the linear polarization produced by a maser is aligned either parallel or perpendicular to the projection of the magnetic field on the plane orthogonal to the direction of propagation of the maser. The polarization of CO suffers a similar ambiguity in magnetic field direction (Goldreich \& Kylafis 1982), so it is possible that the magnetic field direction is in fact parallel to
the large-scale dust lane in DR21. Such a mis-interpretation is very unlikely because, in polarization studies of CO, simultaneous observations of dust polarization have always been used to unwrap this ambiguity (e.g. Lai et al 2003; Vallee \& Fiege 2006). 

Secondly, the polarization of our methanol maser component 1 may have been caused by a \emph{local}
magnetic field (e.g. close to a massive protostar) and this field would be different to
the larger-scale magnetic field across the gas and dust lane. This view is supported by observations of OH polarization in DR21(OH) by Fish et
al. (2005), who found large variations in the direction of the local magnetic field
inferred from the polarization of OH ground-state masers in MM1, although it is possible that these apparent variations are caused by Faraday rotation and not small-scale changes in the magnetic field (Fish \& Reid 2006). If masers show larger variations in linear polarization angle within a region than the dust or CO polarization angles, this suggests that masers (as high-resolution tracers) delineate very localised magnetic field structure. This is clearly very helpful in our attempt to understand the contribution of magnetic fields to individual sites of massive star-formation.
The next step is to determine the polarization properties of OH masers at 1.7-GHz and 6.0-GHz from existing MERLIN data (taking into account differential Faraday rotation), thus providing independent measurements of the magnetic field direction on $\sim$1000 AU (individual maser) scales. Another independent test of the magnetic field in this region will be high-resolution polarimetric dust and CO observations. The magnetic fields inferred from dust and CO polarization are much smoother than those inferred by maser observations, but this may well be an artefact of the much larger beamsizes of the instruments. Dust and/or CO polarization observations at comparable angular resolution to MERLIN should be carried out before conclusions are drawn about the small-scale magnetic fields in the region. 

\subsection{Interior structure of maser spots}

By mapping the positions of the Gaussian-fitted 6.7-GHz methanol maser centroids in each velocity channel, we
have probed the interior structure of individual maser components in regions
significantly smaller than the (50$\times$50 mas) MERLIN beam. High spectral-resolution
studies such as these have previously been carried out using the VLBA at 12.2
GHz by Moscadelli et al. (2003) and at 18-cm by Fish et al. (2006). Despite the much larger beam of the MERLIN observations,
our results are remarkably similar to those from the VLBA, in that we
have been able to map structures such as lines and arcs within regions
smaller than 20 mas. The shape and velocity profile of the structure found with MERLIN must be verified using VLBI, but it is interesting to note that by
fitting Gaussians to individual spectral channels, an `effective angular resolution' (the ability to distinguish the movement of the central peak flux position) an order of magnitude greater than the true angular resolution of MERLIN has been achieved. Follow-up observations have now been made with the European VLBI Network, which will form the basis of a future paper by Harvey-Smith \& Soria-Ruiz.

\section{Conclusions}

By observing the DR21 star-forming complex at 6.7-GHz, two new sites of methanol maser emission in DR21(OH) and
DR21(OH)N have been found. Two dimensional Gaussian components were fit to the
emission in each spectral channel containing a signal greater than
5$\sigma_{RMS}$. The positions of the emission centroids in each channel were then mapped, in an attempt to determine the internal velocity gradients
of the maser spots. This technique added a great deal of information
about the structures traced by the methanol masers, which is vital to
our aim of linking methanol masers to physical features such as discs
around protostars and outflows of circumstellar material. The maser
feature in DR21(OH)N, component 1, is a strong candidate to be a circumstellar disc. It has a double-peaked spectral profile, a complex
velocity gradient that resembles a Keplerian rotation curve and a
magnetic field gradient across its major axis. It also coincides with a
massive protostar that has been previously identified by studies of infra-red and
dust emission. Further study of the existing MERLIN 1.7-GHz and 6.0-GHz OH maser data in full polarizations and high-resolution European VLBI Network data of methanol at 6.7-GHz will hopefully add to our understanding of the disc-like component 1 in particular. Numerical analysis of the feature in terms of a Keplerian rotation model will reveal whether or not this feature is likely to be a disc around a massive protostar.

\section*{Acknowledgments}

The authors thank the referee, Thomas Wilson, for his useful comments. Lisa Harvey-Smith thanks Vincent Fish for very helpful discussions
regarding the AIPS task ORFIT. Ana Duarte-Cabral acknowledges the
support of a JIVE Summer Studentship. MERLIN is a National Facility 
operated by the University of Manchester at Jodrell Bank Observatory 
on behalf of STFC.

\label{lastpage}

\end{document}